\documentclass[preprint]{elsarticle}
\usepackage{amscd,verbatim}
\usepackage[english]{babel}
\usepackage[latin1]{inputenc}
\usepackage{latexsym}
\usepackage{amsmath,amsfonts}
\usepackage{epsf}
\usepackage{hyperref}
\usepackage{natbib,amssymb}
\usepackage{color,graphics}
\journal{Journal of Theoretical Biology}
\begin{document}
\begin{frontmatter}
\title{Deterministic and stochastic aspects of VEGF-A production
and the cooperative behavior of tumoral cell colony}

\author[CSDC,INFN]{Pasquale Laise}
\author[PD]{Francesca Di Patti}
\author[SM]{Duccio Fanelli}
\author[Pat]{Marika Masselli}
\author[Pat]{Annarosa Arcangeli}

\address[CSDC]{CSDC Centro Interdipartimentale per lo Studio di
Dinamiche Complesse, University of Florence, Italy and INFN}
\address[SM]{Dipartimento di Energetica, University of Florence, Via
S. Marta 3, 50139 Florence, Italy}
\address[Pat]{Dipartimento di Patologia e Oncologia Sperimentali, University of
Florence, viale Morgagni 50, 50134 Florence, Italy}
\address[INFN]{INFN - Sezione di Firenze}
\address[PD]{Dipartimento di Fisica G. Galilei,
Universit\`{a} degli Studi di Padova, via Marzolo 8, 35131 Padova, Italy}

\begin{abstract}
A model is  proposed to study the process of
hypoxia-induced angiogenesis in cancer cells. The model accounts for
the role played by the vascular endothelial growth factor (VEGF)-A
in regulating the oxygen intake. VEGF-A is dynamically controlled by
the HIF-1$\alpha$ concentration. If not degraded, HIF-1$\alpha$ can
bind to the subunit termed HIF-1$\beta$ and so experience
translocation to the nucleus, to exert its proper transcriptional
activity. The delicate balance between these opposing tendencies
translates into the emergence of distinct macroscopic behaviors in
terms of the associated molecular concentrations that we here trace
back to normoxia, hypoxia and death regimes. These aspects are firstly
analyzed with reference to the ideal mean-field scenario. Stochastic fluctuations
are also briefly discussed and shown to seed a cooperative interaction
among cellular units, competing for the same oxygen reservoir.
\end{abstract}

\end{frontmatter}
\section{Introduction}\label{sec:intro}
Hypoxia (O$_2$ tension below 2.5 mmHg) is  a hallmark of several
types of solid tumor \cite{tatum}. Growing clinical evidence
postulates a correlation between hypoxia and cancer aggressiveness
and metastasis \cite{koong}. Hypoxia may underlie resistance to
radiotherapy and can modify the cancer phenotype through gene
regulation. A major cellular response  to hypoxia is the
stabilization of hypoxia inducible factor 1 (HIF-1), a transcription
factor that controls cancer progression, and represents a potential
target for therapy \cite{semenza}. HIF-1 is a heterodimer  composed
of a oxygen-dependent HIF-1$\alpha$ and a constitutively expressed
subunit termed HIF-1$\beta$. The transcriptional response to HIF-1
$\alpha$ varies from general effects, such as the up regulation of
anaerobic respiration (virtually in  all tumors), to the more
tissue-specific effects such as angiogenesis \cite{tatum}. The
tumor-driven angiogenesis is therefore a process pivotal in tumor
growth and progression, and is mainly related to the HIF-1 dependent
secretion of the angiogenic factor Vascular Endotelial Growth Factor
-A (VEGF-A) \cite{vaupel}. Low ${O}_2$ tension may shape the cancer
phenotype. According to a recently proposed model for cancer
progression \cite{fang}, the tumor milieu exerts Darwinian selection
in favor of cancer cells. The basis for this selection can be
two-fold: firstly, the milieu may activate tumor-specific pathways.
A second basis for selection may reside in the ability of cells to
protect the intracellular compartment from a hypoxic milieu. Given
the important role of hypoxia in cancer progression \cite{fang},
targeting the cellular responses to hypoxia, including the
triggering of VEGF-A secretion and the ensuing angiogenesis, may
form an alternative to, or a component of, the current
chemotherapeutical practice.

In this paper, a  model is proposed to resolve the complex dynamical
interplay between the molecular chemical species that participate to
the aforementioned process. In particular, we will start by
exploring the mean-field scenario stemming from the hypothesized
networks of interaction. This formally amounts to operate in the
limit for infinite system size, the species concentrations obeying
to a closed set of ordinary differential equations. We target our
analysis to the asymptotic (equilibrium) regime highlighting the
emergence of different attractive (stable) states of the single cell
dynamics, which appear to be selectively chosen depending on the
chemical parameters involved. We then turn to consider the evolution
of three neighbors cells which are effectively coupled via the
external environment, and adjust consequently their interior
dynamics. The role of fluctuations stemming from the discreteness of
the system under scrutiny, is  shortly addressed resorting to direct
stochastic simulations. Depending on the selected parameters,
normoxic condition can develop as an emergent dynamical equilibrium.
Allowing for punctual mutations, so to mimic tumor derive, can alter
the stability of the system, and eventually result in a competition
between distinct cell populations. Indeed, as we shall demonstrate,
cooperative mechanisms  arise being mediated by the stochastic
component of the dynamics. Interestingly, adjacent cells can also
profit from the gained ability of an individual cell unit to survive
under hypoxic condition, an effect which fades off in the continuum
limit.

The paper is organized as follows.  Next section is devoted to
introducing the key ingredients of the model, here formulated as an
ensemble of chemical equations.  Then the mean-field version of the
model is studied, which in turn corresponds to work in the infinite
system size limit. The focus is on a single cell: different
dynamical regimes are identified corresponding to distinct choices
of the chemical parameters. Stochastic simulations are also
performed to test the adequacy of the theory predictions and
challenge in silico the role played by finite size corrections. Then
in Sec. \ref{sec:three_cells} we turn to discussing a generalization
of the proposed model where three cells are made to interact and
compete for the oxygen amount. Finally, in Sec.
\ref{sec:conclusions}, we sum up and draw our conclusions.

\section{A chemical model of hypoxia}
Consider now an individual cell and  assume the membrane barrier to
divide between interior and exterior regions. Selected molecules can
migrate through the semi-permeable membrane, being hence transported
from the outside to the inside and viceversa. Aiming at elaborating
a sound description of the hypoxia cycle \cite{vaupel}, we select a
set of candidate molecules which define a close ensemble of
interacting elements, according to our schematic representation of
the phenomenon under scrutiny. Oxygen molecules ($\text{O}_2$)
populate the external milieu. They diffuse and eventually reach the
membrane where they happen to combine with the hydroxylases
($\text{W}$), which are consequently turned into an active phase,
hereafter labeled $\text{W}_a$\footnote{Hydroxylases are
intracellular enzymes, organized in three families respectively
PHD1, PHD2, PHD3. Hydroxylases hence populate the internal flow and
eventually combine with the oxygen that have penetrated the cell. As
formalized in the following, we shall here focus on a more simple
picture, and imagine that the encounters between hydroxylases and
oxygen occur at cell boundary. In such a way, hydroxylases can be
ideally pictured as antennae, localized on the membrane wall,
chasing for the (outer) oxygen molecules.}. This process is
encapsulated into the following chemical equation:
\begin{equation}\label{eq:chem1}
\text{O}_2+ \text{W}   \stackrel{a}{\longrightarrow }  \text{W}_a +
\text{E}_0
\end{equation}
where $a$ is  the reaction rate and $\text{E}_0$ stands for the
empty case which the oxygen leaves behind when it gets absorbed by
the $\text{W}$. The label $0$ refers to the chemicals which are
populating the outside region, while the molecules confined inside
the cell wall are targeted with the subscript $i$. By introducing
the concept of empty spaces (E$_0$ outside, and E$_i$ inside) we
impose that the total number of elements (including the empties) is
a constant of the dynamics \cite{mckane,dipatti}. Active
hydroxylases $\text{W}_a$ interfere with the $\text{HIF-}\alpha$,
yielding to the degradation via the proteasome pathway:
\begin{equation}\label{eq:chem2}
\text{HIF-}\alpha + \text{W}_a   \stackrel{b}{\longrightarrow }
\text{W} + \text{E}_i
\end{equation}

Once the $\text{HIF-}\alpha$ is being targeted to degradation, the
hydroxylases go back to the primitive configuration, waiting for the
next oxygen to drive the transition into its active state. The
degraded $\text{HIF-}\alpha$ molecule is replaced by the empty inner
element $\text{E}_i$.

Moreover, the $\text{HIF-}\alpha$ can occasionally meet the
$\text{HIF-}\beta$ subunit. This encounter necessarily takes place
within the cell and gives rise to the $\text{VEGF}_i$
macromolecule\footnote{From hereon we will simply refer to VEGF-A as
to VEGF.}. The corresponding chemical equation reads:

\begin{equation}\label{eq:chem3}
\text{HIF-}\alpha +\text{HIF-}\beta   \stackrel{c}{\longrightarrow}
 \text{VEGF}_i + \text{E}_i
\end{equation}

$\text{VEGF}_i$ can eventually abandon  the cell interior to occupy
an external vacancy $\text{E}_0$, which is hence mutated into a
$\text{VEGF}_0$ element. In formulae:

\begin{equation}\label{eq:chem4}
\text{VEGF}_i + \text{E}_0   \stackrel{e}{\longrightarrow }
\text{VEGF}_0 + \text{E}_i
\end{equation}

Furthermore, the $\text{VEGF}_0$ participates to the  network of
extracellular reactions which underlay the angiogenesis process. In
practice, following a complex cascade of nested chemical reactions,
the $\text{VEGF}_0$ attracts the oxygen $\text{O}_2$, which is hence
imagined to replace an (external) empty case. This materializes in
turn into a simplified vision of a complex dynamic pathway, which
has the sole scope of providing a self-consistent entry to the
inspected process. The corresponding chemical equation takes the
form:

\begin{equation}\label{eq:chem5}
\text{VEGF}_0 + \text{E}_0   \stackrel{f}{\longrightarrow }
\text{VEGF}_0 + \text{O}_2
\end{equation}

Notice that the $\text{VEGF}_0$ molecule is also a product  of the
reaction, a working hypothesis that we motivate with the empirical
observation that $\text{VEGF}_0$ can attract several oxygen units.
This process will eventually come to a stop, as follows
$\text{VEGF}$ degradation which occurs both inside and outside the
cells:

\begin{eqnarray}
\text{VEGF}_0 &  \stackrel{d}{\longrightarrow } & \text{E}_0 \label{eq:chem6}\\
\text{VEGF}_i &  \stackrel{d}{\longrightarrow } & \text{E}_i
\label{eq:chem7}
\end{eqnarray}

Finally, we should  also account for the constitutive generation of
$\text{HIF-}\alpha$ and $\text{HIF-}\beta$:
\begin{eqnarray}
\text{E}_i &  \stackrel{g_\alpha}{\longrightarrow } & \text{HIF-}\alpha \label{eq:chem8}\\
\text{E}_i &  \stackrel{g_\beta}{\longrightarrow } &
\text{HIF-}\beta \label{eq:chem9}
\end{eqnarray}

\begin{figure}[tb]
\begin{center}
\includegraphics[width=8cm]{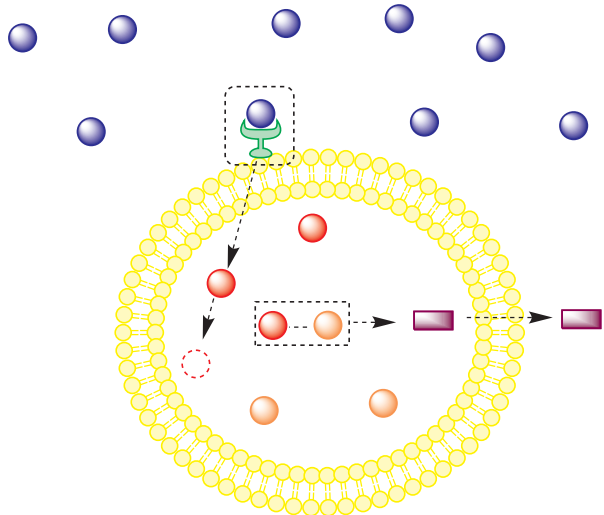}
\caption{Cartoon for a single  cell dynamics as follows the proposed
scheme. Oxygen molecules (circles laying outside the cell) are
captured by the hydroxylases sitting on the membrane, and so trigger
the HIF-$\alpha$ (dark circles, red online, stored inside the
membrane) elimination pathways (dashed circle). At the same  time,
HIF-$\alpha$ can combine with HIF-$\beta$ (light circles, orange
online, stored inside the membrane), yielding to the (barrel like,
in the picture) VEGF$_i$. This latter can exit the cell wall, so
ensuing in VEGF$_0$ and drive the incoming oxygen
flux.\label{fig:cartoon}}
\end{center}
\end{figure}

The scenario  outlined above is depicted in Fig. \ref{fig:cartoon},
where the main molecular actors are also
specified\footnote{Importantly, the oxygen is initially imagined to
fill the external reservoir, and solely re-integrated via the VEGF
pathway. This is clearly a simplifying assumption, as in general one
should also account for the oxygen supply via blood flux. On the
other hand, we are here interested in the ability of the cell to
self-organize when exposed to stressing condition, as it is the lack
of oxygen income, a working hypothesis  which justifies the assumed
scenario.}. It should be noted that the above model conserves the
number of molecules which totals in $N$, a fact that is ultimately
related to the presence of the empties species $E$.

We should emphasize that the proposed model  is intimately
stochastic, in its original chemical formulation. The  inherent
stochasticity emerges as an effect of the discreteness of the
medium: finite size corrections, also termed demographic noise,
result in a endogenous perturbation which can significantly impact
the dynamics of the system as compared to the idealized continuum
picture, which formally applies in the limit of infinite
constituents. To respect the stochastic nature of the problem and
simulate it in silico, one can resort to the celebrated Gillespie
algorithm \cite{gillespie}, a method which makes it possible to
\emph{exactly} monitor the time evolution of the \emph{discrete}
population of mutually interacting entities. The scheme captures in
fact the probabilistic nature of the microscopic couplings and
enables one to resolve the contribution associated to finite size
fluctuations. As we shall demonstrate, stochastic fluctuations can
eventually materialize in complex dynamics, which call for sound
biological interpretation.

As anticipated, and opposed to the probabilistic approach, one can
invoke the deterministic approximation, which in turn corresponds to
assuming an infinite population  amount and consequently
disregarding finite size correlations. This defines the mean-field,
continuum level of description, to which next sections are entirely
devoted. More specifically, we will start by considering the case of
a single cell, which we will discuss with reference to its mean
field analogue. Notice that the deterministic formulation results
from a straightforward derivation, that moves from the underlying
master equation for the stochastic process, as defined by the above
set of chemical equations. In the following, we will allude to such
procedure, recalling its general philosophy, and just provide the
end result for the case under inspection, without insisting on the
technicalities.  Working  within the continuum representation we
will then elaborate on the asymptotic dynamics of the recovered
system of ordinary differential equations, by identifying distinct
fixed points and discussing their associated stability
characteristics. The inherent stochastic effects stemming from
finite $N$ corrections, will be relegated to section 5.

\section{Master equation}

At time $t$, the status of the chemical system, as ruled by the
above reactions, is completely specified once the species amount are
assigned. These are discrete entries, one for each species belonging to
the ensemble  of interacting population, which define the scalar
components of the status vector $\mathbf{n}$. Recall that the system
is intrinsically stochastic, its evolution being a suite of events
which realizes with a prescribed probability. In mathematical terms,
this fact implies dealing with the probability that the system is
seen in the state $\mathbf{n}$ at time $t$, hereafter denoted
$P(\mathbf{n},t)$. This latter quantity adjusts in time and obeys to
the master equation:
\begin{equation}\label{eq:master}
\frac{d}{dt}P(\mathbf{n},t)= \sum_{\mathbf{n}'} \left [
T(\mathbf{n}|\mathbf{n}') P(\mathbf{n}',t) -
T(\mathbf{n}'|\mathbf{n}) P(\mathbf{n},t)\right]
\end{equation}
where $T(\mathbf{n}'|\mathbf{n})$ denotes the transition rates
between two adjacent states, from $\mathbf{n}$ to $\mathbf{n}'$,
compatible with the chemical constraints imposed by equations
(\ref{eq:chem1})-(\ref{eq:chem9}).  For the sake of compactness, we
omit here to list the specific forms of the transition rates
relative to the model under analysis.

The partial differential equation  (\ref{eq:master}) is an exact
formulation, which can be in principle analyzed to track the system
dynamics. In practice however, it is hard to analytically handle and
further progress has to rely on direct numerical investigations.

The Gillespie algorithm  represents in particular a viable strategy
to simulate the time evolution of the probability distribution
$P(\mathbf{n},t)$. The method sets up a Monte Carlo procedure to
determine the next reaction that is selected to occur, among all
possible ones, and  the time $\tau$, when the event takes place.
Both selections reflect the assigned transition probabilities, which
scales proportional to the number of substrate molecules times the
associated reaction constants. Then, based on the selected reaction,
the update of the molecule count follows, and time $t$ is replaced
by its value increased by the stochastic time interval $\tau$,
namely $t \rightarrow t+\tau$.

Alternatively, the master equation (\ref{eq:master}) can be studied
via perturbative methods, aimed at extracting both the idealized
continuum picture and the successive finite size corrections. The
van Kampen  large $N$ expansion is in this respect a widely adopted
technique, extensively characterized in the literature
\cite{Gardiner, vanKampen}. In the following section, we shall
operate within the van Kampen framework to derive a closed system of
ordinary differential equations for the coupled evolution of the
concentrations of chemical system (\ref{eq:chem1})-(\ref{eq:chem9})
in the limit of large (formally infinite) $N$.

\section{Mean-field solution of the single cell model}
Let us focus on a single cell, as pictorially  represented in Fig.
\ref{fig:cartoon}. Consider  for instance the O$_2$ species.
According to van Kampen \cite{vanKampen}, one can put forward the
following ansatz for the associated mean-field (continuous)
concentration here labeled $\mathcal{O}$:

\begin{equation*}
\mathcal{O} = \frac{n_{O_2}}{N} + \frac{\xi}{\sqrt{N}}
\end{equation*}
where the extensive quantity $n_{O_2}$  is  discrete: it is one of
the components of the vector $\mathbf{n}$ and stands for the number
of oxygen molecules. $\xi$ is instead a stochastic variable and
contributes with an additive terms which scales as $1/\sqrt{N}$, so
materializing in a perturbation to the idealized mean-field
dynamics. Stochastic effects can be hence safely neglected when
performing the continuum limit \cite{Gardiner, vanKampen}, the relic
concentration obeying to a deterministic set  of coupled
differential equations. We recall however that beyond the idealized
scenario,  finite size corrections do matter when real systems are
concerned, the intimate discreteness  being a crucial ingredient of
the  dynamics. We shall return on this issue in Section 5, to
highlight the important impact of such an endogenous source of
stochastic perturbation with reference to the scrutinized model.

For simplicity, by extending  the above reasoning to the other
involved species, we will here denote by $\mathcal{W}$,
$\mathcal{W}_a$,  $\mathcal{V}_0$, $\mathcal{V}_{i}$,
$\mathcal{E}_{0}$, $\mathcal{E}_{i}$, $\mathcal{H}_{\alpha}$ and
$\mathcal{H}_{\beta}$ the average (mean-field) concentration of W,
W$_a$, VEGF$_0$, VEGF$_i$, E$_0$, $E_i$, HIF-$\alpha$ and
HIF-$\beta$ respectively. According to this notation, the mean-field
equations take the form\footnote{In deriving eqs. (\ref{eq:mf8})
correlations have been neglected, a legitimate choice in the
mean-field approximation and whose validity has been tested versus
numerical simulation in a number of models that follow the same
conceptual scheme \cite{alan2,fra2}.}:
\begin{eqnarray}
\frac{d}{dt}\mathcal{O} & = & -a \mathcal{O} \mathcal{W} +
f \mathcal{V}_0 \mathcal{E}_{0}  \label{eq:mf1} \\
\frac{d}{dt}\mathcal{W} & = & -a \mathcal{O} \mathcal{W}
+b \mathcal{W}_a \mathcal{H}_{\alpha} \label{eq:mf2} \\
\frac{d}{dt}\mathcal{W}_a & = & a \mathcal{O} \mathcal{W} -
b \mathcal{W}_a \mathcal{H}_{\alpha} \label{eq:mf3} \\
\frac{d}{dt}\mathcal{H}_{\alpha} & = & - b \mathcal{W}_a \mathcal{H}_{\alpha}
 -c \mathcal{H}_{\alpha} \mathcal{H}_{\beta} +g_\alpha \mathcal{E}_{i} \label{eq:mf4}\\
\frac{d}{dt}\mathcal{H}_{\beta} & = & -c \mathcal{H}_{\alpha}
\mathcal{H}_{\beta}
+g_\beta \mathcal{E}_{i}  \label{eq:mf5} \\
\frac{d}{dt}\mathcal{V}_{i} & = & c \mathcal{H}_{\alpha}
\mathcal{H}_{\beta}
- e \mathcal{V}_{i} \mathcal{E}_0 -d \mathcal{V}_{i} \label{eq:mf6} \\
\frac{d}{dt}\mathcal{V}_0 & = & e \mathcal{V}_{i} \mathcal{E}_{0}  - d \mathcal{V}_0  \label{eq:mf7} \\
\frac{d}{dt}\mathcal{E}_{i} & = & b \mathcal{W}_a
\mathcal{H}_{\alpha} + c \mathcal{H}_{\alpha} \mathcal{H}_{\beta} +
e \mathcal{E}_{0}  \mathcal{V}_{i} + d \mathcal{V}_{i} -
(g_\alpha+g_\beta) \mathcal{E}_{i} \label{eq:mf8}
\end{eqnarray}
with the conservation law
\begin{equation}\label{eq:concentration_law}
\mathcal{O}+ \mathcal{W} + \mathcal{V}_0 + \mathcal{V}_{i} +
\mathcal{E}_{0} + \mathcal{W}_a + \mathcal{E}_{i} +
\mathcal{H}_{\alpha} + \mathcal{H}_{\beta} =1
\end{equation}

Notice that the same result could be recovered without the use of
the van Kampen's expansion. Indeed, multiplying both sides of the
master equation by a scalar component of the status vector
$\mathbf{n}$, summing over all $\mathbf{n}$, shifting some of the
involved sums by $\pm 1$ and neglecting the correlations (which is
legitimate in the $N$ infinite limit) one recovers the same  mean
field system.

From Eq. (\ref{eq:mf2}) and  (\ref{eq:mf3}), it straightforwardly
follows that $\mathcal{W}_a + \mathcal{W} $ stays constant: This
invariant quantity is hereafter set equal to $n_0$ and specified by
the initial condition. To investigate the dynamics of the above
system (\ref{eq:mf1}) - (\ref{eq:mf8}), and in particular elaborate
on its asymptotic evolution, one can look for stationary solutions.
These are the fixed points of the dynamics, which can be explicitly
found by setting to zero the derivatives in (\ref{eq:mf1}) -
(\ref{eq:mf8}) and solving the obtained algebraic system, where the
(final) concentration amounts enter as unknowns.
\subsection{Stationary states}
Solving for the fixed points  returns three possible solutions.
These latter are here characterized and labeled via the progressive
indexes $i=1,2,3$, associated to the predicted asymptotic
concentrations. Two solutions can be readily obtained via direct
inspection of the above equations, and correspond to opposite
tendencies that we will hereafter call {\it normoxia} and {\it
death} respectively.

Normoxia conditions are assumed to correspond to the asymptotic
solution:
\begin{equation}
\mathcal{V}_{0,1}  =
\mathcal{V}_{i,1}=\mathcal{H}_{\alpha,1}=\mathcal{W}_1=\mathcal{E}_{i,1}=0.
\end{equation}
HIF-$\alpha$ is  turned off and so are the inner/outer
concentrations of VEGF. The HIF-$\alpha$ degradation mechanism hence
prevails over the alternative pathway that yields VEGF production,
and which could in principle contribute to enhance the oxygen
supply. The amount of O$_2$, here quantified by $\mathcal{O}_1$, is
on the contrary sufficient to guarantee the correct cell
functioning. Hydroxylases are frozen in their active state W$_a$ and
consequently stimulate the repression of the HIF-$\alpha$
quota\footnote{In the transient dynamics, HIF$-\alpha$ molecules are
indeed present and translate into VEGF entities, upon combination
with HIF$-\beta$. VEGF molecules can remain temporarily active also
when the HIF$-\alpha$ has been eliminated, and so drive a consequent
oxygen influx. There is hence a certain degree of inertia which
increments the initial oxygen amount. The effect becomes more
pronounced as $g_\alpha$ gets larger, more HIF$-\alpha$ populating
the cell during the transient dynamics, before degradation has
occurred. However, provided $g_\alpha$ is small enough, the cell
reaches a steady state configuration, with no residual amount of
HIF$-\alpha$ molecules. For this reason, we refer to this
configuration as to the normoxia regime.}. From the conservation
relations we immediately deduce the following relation:
\begin{eqnarray}
\mathcal{W}_{a,1} &=& n_0 \\
\mathcal{O}_1 + \mathcal{E}_{0,1} + \mathcal{H}_{\beta,1} &=& 1-n_0
\end{eqnarray}
The outer cases  are shared between oxygen and empties, as
exemplified by the latter equation, in a proportion that cannot be
predicted a priori.

As opposed to  normoxia condition, a trivial solution of the mean
field problem is found when:
\begin{equation}
\mathcal{W}_{a,3}=\mathcal{V}_{0,3}  =
\mathcal{V}_{i,3}=\mathcal{H}_{\alpha,3}=\mathcal{E}_{i,3}=\mathcal{O}_3
= 0
\end{equation}
with in  addition $\mathcal{W}_3=n_0$ and $\mathcal{E}_{0,3} +
\mathcal{H}_{\beta,3}=1-n_0$. In this case, the cell is dead: the
oxygen amount is zero and the hydroxylases cycle is not active, so
denouncing the lethargic state of the cell device.

An intermediate asymptotic stationary state is also  possible, when
the cell balances the oxygen consumption by properly adjusting the
VEGF production cycle, at equilibrium.  Upon manipulation of the
fixed point equations associated to the system (\ref{eq:mf1}) -
(\ref{eq:mf8}), the  equilibrium solution for $\mathcal{E}_{0,2}$ is
found to verify the following quadratic equation:
\begin{equation}\label{eq:equation_for_m_bar}
   \frac{f e}{d} \mathcal{E}_{0,2}^2 -\left ( \frac{g_\alpha}{g_\beta} -1\right ) d \mathcal{E}_{0,2}
   -d \left ( \frac{g_\alpha}{g_\beta} -1 \right )=0
\end{equation}

The previous equation admits a solution only if  $g_\alpha>g_\beta$,
otherwise is reduced to a sum of three positive terms which cannot
mutually cancel for positive coefficients and  state variables. When
$g_\alpha=g_\beta$, then $\mathcal{E}_{0,2}=0$, which corresponds to
having the largest allowed oxygen concentration $\mathcal{O}_2 =
1-n_0$. This latter can alternatively be seen as a limiting normoxia
condition. All externally available cases are simultaneously
occupied by oxygen molecules, which can still effectively contrast
the HIF-$\alpha$ generation, while providing the necessary supply
for the cell functioning ($\mathcal{W}_{a,1} \ne 0$). Beyond this
condition, a quota of HIF-$\alpha$ survives the degradation and
sustains the VEGF production process. VEGF molecules transported
outside the cell support the afflux of  oxygen and so contribute to
stabilize the quota of HIF-$\alpha$. In practical terms, the
condition $g_\alpha=g_\beta$ (when the HIF$-\alpha$ and HIF$-\beta$
production rates equalize) signals a transition between two distinct
asymptotic regimes, a fact that could be further appreciated by
quantifying the values assumed by the remaining involved
concentrations. We identify the intermediate fixed point with the
{\it hypoxia} conditions: The cell is in fact alive thanks to a
dedicated feedback that integrates the original oxygen amount via
the VEGF production. Once $g_\alpha>g_\beta$ the solution of Eq.
(\ref{eq:equation_for_m_bar}) reads:
\begin{equation}\label{eq:m_bar}
\mathcal{E}_{0,2}= \frac{d\left [ e(g_\alpha-g_\beta) +\displaystyle
\sqrt{e(g_\alpha-g_\beta)\left[ e(g_\alpha-g_\beta)+4fg_\beta)
\right]}\right ]}{2 f e g_\beta }
\end{equation}

The equilibrium values of the other variables cannot  be univocally
determined. However, given the value of $\mathcal{E}_{0,2}$ as
estimated above, the following ratios are fixed:
\begin{eqnarray*}
\frac{\mathcal{O}_2 \mathcal{W}_2}{\mathcal{V}_{0,2} }   & = & \frac{f}{a} \mathcal{E}_{0,2} \\
\frac{\mathcal{W}_{a,2} \mathcal{H}_{\alpha,2}}{\mathcal{V}_{0,2}} & = & \frac{f}{b} \mathcal{E}_{0,2} \\
\frac{\mathcal{H}_{\alpha,2}
\mathcal{H}_{\beta,2}}{\mathcal{V}_{0,2} } & =
& \frac{g_\beta}{c} \frac{f}{g_\alpha - g_\beta} \mathcal{E}_{0,2}\\
\frac{\mathcal{E}_{i,2} }{\mathcal{V}_{0,2} }& = & \frac{f}{g_\alpha - g_\beta} \mathcal{E}_{0,2} \\
\frac{\mathcal{V}_{i,2}}{\mathcal{V}_{0,2}} & = & \frac{d}{e}
\frac{1}{\mathcal{E}_{0,2}}
\end{eqnarray*}

\begin{figure}[tb]
\begin{center}
\includegraphics[width=8cm]{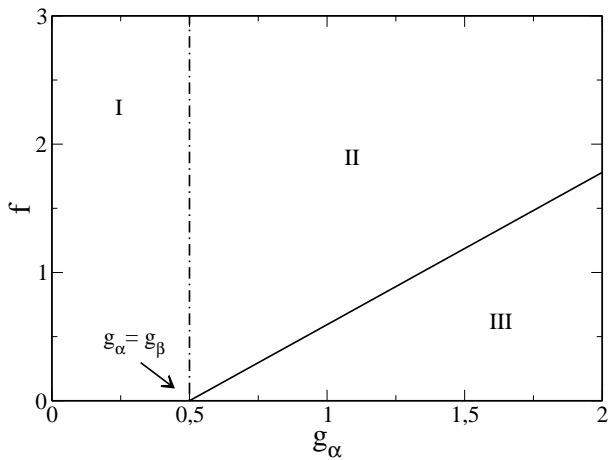}
\caption{The existence  of the normoxia (I), hypoxia (II) and death
(III) regions is displayed in the $(g_\alpha,f)$ plane, as predicted
by the mean-field calculations for $N=1000$, $n_0=20$, $e=0.9$,
$d=0.4$ and $g_\beta=0.5$. The transition from I to II is found to
occur when $g_\alpha=g_\beta$, i.e. when the HIF$-\alpha$ production
rate prevails over the HIF$-\beta$ one. Then the exceeding
HIF$-\alpha$ is stabilized to an asymptotic fixed concentration via
the additional VEGF pathway. It can be further appreciated that
region II gets compressed when reducing the control parameter $f$.
\label{fig:plane_e_g}}
\end{center}
\end{figure}

Notice that an implicit  relation between $\mathcal{W}_{a,2}$ and
$\mathcal{V}_{0,2}$ is also derived by making  use of Eq.
(\ref{eq:concentration_law}). More interestingly, for the case
$g_\alpha > g_\beta $, $\mathcal{E}_{0,2}$ can be shown to increase
with $g_\alpha$, while keeping all the other parameters constant.
The maximum value that can  be eventually attained is
$\mathcal{E}_{0,max} = 1-n_0$, a limiting condition that, together
with $g_\alpha>g_\beta$, identifies the portion of parameters plane
for which the hypoxia fixed point do exist. More specifically, the
largest the values of $f$, i.e. the rate of oxygen production via
the VEGF cycle, the wider the window in $g_\alpha$ that delimits the
hypoxia regime. Working in the $(g_\alpha,f)$ plane we can
analytically determine the boundaries of the domain of existence of
the non trivial fixed point. Its equation can be deduced by imposing
$\mathcal{E}_{0,2}=\mathcal{E}_{0,max}$ into Eq.
(\ref{eq:equation_for_m_bar}) which immediately yields to
\begin{equation*}
f = \frac{e d \mathcal{E}_{0,max} + d^2}{ g_\beta e
\mathcal{E}_{0,max}^2} g_\alpha - \frac{e d \mathcal{E}_{0,max} +
d^2}{e \mathcal{E}_{0,max}^2}
\end{equation*}
This is a line which  crosses the axis $f=0$ when
$g_\alpha=g_\beta$, see Fig. \ref{fig:plane_e_g} where three
domains, respectively I, II, III, are outlined. Notice  that the
normoxia and death solutions can in principle extend over all the
parameters plane, while the hypoxia is solely bound to region II. The
hypoxia region shrinks as $f$ is reduced and eventually fades off
when the VEGF's ability to carry oxygen is hindered. We recall in
fact that within our simplified formulation, the oxygen income is
ultimately controlled by the parameter $f$, and consequently linked
to the available $\text{VEGF}_0$ quota. It is hence plausible that
that region here identified to correspond to the hypoxia regime
vanishes when  $f$ is set to zero. Numerical simulations indicate
that the basin of attraction of the normoxia state prevails in
region I, while the death configuration dynamically takes over in
region III. In region II, it is the hypoxia condition to represent
the stable attractor of the dynamics. For this reason, region I, II
and III are respectively termed normoxia, hypoxia and death, and the
caption of Fig. \ref{fig:plane_e_g} reflects such a choice. To gain
further insight into this empirical observation, we turn in the next
section to studying the stability properties of the identified fixed
point. The analysis is carried out via combined numerical and
analytical means, as the complexity of the model prevents us from
performing a complete analytical treatment.

\subsection{Stability analysis}

In this section we set down to study the stability properties of the
three families of equilibrium points, as identified above. To this
end  we calculate the Jacobian matrix  associated to the system
(\ref{eq:mf1})-(\ref{eq:mf8}), explicitly given in the Appendix and
evaluate the eigenvalues relative to the three fixed points, namely
the normoxia, the hypoxia and the death.

Consider first the solution that corresponds to the cellular death,
previously labeled with the index $3$ in the concentration. We
recall that $\mathcal{W}_{a,3}=\mathcal{V}_{0,3} =
\mathcal{V}_{i,3}=\mathcal{H}_{\alpha,3}=\mathcal{E}_{i,3}=\mathcal{O}_3
= 0$, $\mathcal{W}_3=n_0$, while the concentration
$\mathcal{E}_{0,3}$ and $\mathcal{H}_{\beta,3}$ can in principle
take any value, constrained by the relation $\mathcal{E}_{0,3} +
\mathcal{H}_{\beta,3}=1-n_0$, which ultimately specifies the family
of possible solutions. The associated characteristic polynomial
reads:
\begin{equation}\label{eq:pol3}
 \lambda^3 (\lambda + a n_0 ) (\lambda +d)
\det \left( \begin{array}{ccc}
 \lambda +c \mathcal{H}_{\beta,3}        & 0 & -g_\alpha \\
 -c \mathcal{H}_{\beta,3}        & \lambda  +e \mathcal{E}_{0,3} +d    & 0 \\
 -c \mathcal{H}_{\beta,3} & - e \mathcal{E}_{0,3} - d & \lambda +g_\alpha +g_\beta
\end{array} \right) =0
\end{equation}
where $\lambda$ stands for the eigenvalue. It is evident that
$\lambda_1=\lambda_2=\lambda_3 =0$, $\lambda_4=-a n_0$ and
$\lambda_5=-d$ are possible solution of Eq. (\ref{eq:pol3}). The
remaining three eigenvalues cannot be analytically determined. For
$g_\alpha> g_\beta$ (a condition that embraces region II and III of
the parameters plane depicted in Fig. \ref{fig:plane_e_g}),
analytical progress is possible: just one stable solution exists
among those  of type $3$ that are in principle admissible. To
substantiate this observation, let us denote with $J_3^{sub}$ the
non trivial sub-matrix of the full Jacobian (\ref{eq:pol3}):
\begin{displaymath}
J_3^{sub} = \left( \begin{array}{ccc}
 -c \mathcal{H}_{\beta,3}        & 0 & +g_\alpha \\
 +c \mathcal{H}_{\beta,3}        & -e \mathcal{E}_{0,3} -d    & 0 \\
 +c \mathcal{H}_{\beta,3} & +e \mathcal{E}_{0,3} + d & -g_\alpha -g_\beta
\end{array} \right).
\end{displaymath}
It is easy to see that $\det(J_3^{sub})=c  \mathcal{H}_{\beta}
(g_\alpha-g_\beta) (d+e\mathcal{E}_{0,3})$, and so $\det
(J_3^{sub})>0$ $\forall \mathcal{H}_{\beta,3} >0$ and
$g_\alpha>g_\beta$. Positive determinant implies unstable solutions,
as can be immediately appreciated with simple algebraic
considerations. This means that the fixed point corresponding to the
death state is always unstable when $g_\alpha>g_\beta$, and
$\mathcal{H}_{\beta,3} \ne 0$, this latter quantity being positively
defined. At variance, stable solution can manifest in the region
specified by condition $g_\alpha>g_\beta$, when
$\mathcal{H}_{\beta,3}=0$. This latter request equivalently implies
$\mathcal{E}_{0,3} = 1-n_0$, and removes the degeneracy (at least in
region I and II) by uniquely characterizing the fixed point of type
III selected by the dynamics. The only stable fixed point of type
``death'' in region II and III, is now completely specified and the
three lacking eigenvalues can be readily calculated, resulting in
$\lambda_6=0$, $\lambda_7=-e (1-n_0) -d $ and $\lambda_8=-g_\alpha
-g_\beta$. Solutions of type $3$ are clearly stable also in region
I, suggesting that the cell can always meet the conditions that
drive its death, an asymptotic fate which is ultimately determined
by the selected initial datum (corresponding to the environmental
conditions).

However, direct numerical inspection, suggests that within region
II, the system tends to converge to the hypoxia solution, this
latter being hence dynamically favored over the stable death regime.
Tracing the respective basins of attraction can be tackled via a
dedicated campaign of numerical simulations. Biologically, it can be
hence speculated that for sufficiently large values of the parameter
$g_\alpha$ the cell opposes the death derive, by compressing its
respective basin of attraction upon activation of the hypoxia cycle.

In region I,  when $g_\alpha<g_\beta$, an infinite family of
solutions of type $3$ can in principle manifest, depending on the
specific initial datum. The associated eigenvalues can be calculated
numerically by tuning both $\mathcal{H}_{\beta,3}$ and
$\mathcal{E}_{0,3}$ within the allowed range of variation, pointing
to the stability of the solutions. Again, the cell can always die in
response to particularly stressing conditions. To complete the
description of region I, one needs to address the stability
properties of the normoxia solution. In this case, repeating the
above strategy, the characteristic polynomial turns out to be
\begin{displaymath}
\lambda^3 (\lambda + a \mathcal{O}_1 ) (\lambda +d) \det \left(
\begin{array}{ccc}
 \lambda +b \mathcal{W}_{a,1} +c \mathcal{H}_{\beta,1}        & 0 & -g_\alpha \\
 -c \mathcal{H}_{\beta,1}        & \lambda  +e \mathcal{E}_{0,1} +d    & 0 \\
 -b \mathcal{W}_{a,1} -c \mathcal{H}_{\beta,1} & - e \mathcal{E}_{0,1} - d & \lambda +g_\alpha +g_\beta
\end{array} \right) =0
\end{displaymath}
which immediately gives $\lambda_1=\lambda_2=\lambda_3 =0$,
$\lambda_4=-a\mathcal{O}_1$ and $\lambda_5=-d$. In this case the
determinant of the residual $3 \times 3$ sub-matrix does not return
any sensible information on the sign of the real part of the three
unknown eigenvalues. To overcome this limitation we proceed as
follows: we first integrate numerically the governing system of
differential equations and then select the attained equilibrium
fixed point,  as an entry to calculate the full set of eigenvalues.
Among the missing eigenvalues, two (which can be complex or real
depending on $g_\alpha$) always display a negative real part, while
the third one is  real and can change the sign as $g_\alpha$ is
increased. This situation is illustrated in the inset of Fig.
\ref{transition_num}: the real eigenvalue is negative over a range
of parameters that practically corresponds to region I, and it
approaches zero for $g_\alpha \gtrsim g_\beta$. This result holds
true for different parameters choice (i.e. the value of $f$) and
selected initial conditions. It can hence be argued that  the
normoxia condition is stable within region I and that it gets
unstable when the boundary of region II, the domain of existence of
the hypoxia solution, is (continuously) approached. Indeed, within
region I, the competition between the two attractors of type $1$
(normoxia) and $3$ (death) is dynamically resolved to favor the
former, as confirmed by direct numerical inspection.

In conclusion, the proposed model can alternatively yield to three
distinct dynamical regimes, depending on the specific choice of the
chemical parameters. The normoxia condition is stable in region I
and loses its stability when the edges of the domain II are
approached. In this latter region it is the hypoxia solution to take
over, by prevailing over the death regime which instead dominates in
region III. Interestingly,  death is always an attractive state of
the dynamics, irrespectively of the selected parameters, a tendency
that the cell  efficiently opposes in region I and II. Formally, it
is hence not accurate to talk about a transition among the three
states, the landscape of possible stable attractors being more
complex as outlined above. However, to keep the message simple, we
shall hereafter reflect the practical segregation into three
distinct zones, as virtual transitions. Notice that a punctual
mutation of the reference parameters could substantially alter the
cell functioning and determine its death.  In the next subsection we
turn to numerical simulations to validate the proposed scenario. The
simulations here performed refer both to the idealized mean-field
dynamics and to the stochastic scenario which respects the inherent
discreteness of the chemical medium.

\subsection{Numerical simulations vs. mean field
theory} 
\begin{figure}[tb]
\begin {center}
\begin {tabular}{cc}
\includegraphics[width=5.5cm, height=3.2 cm ]{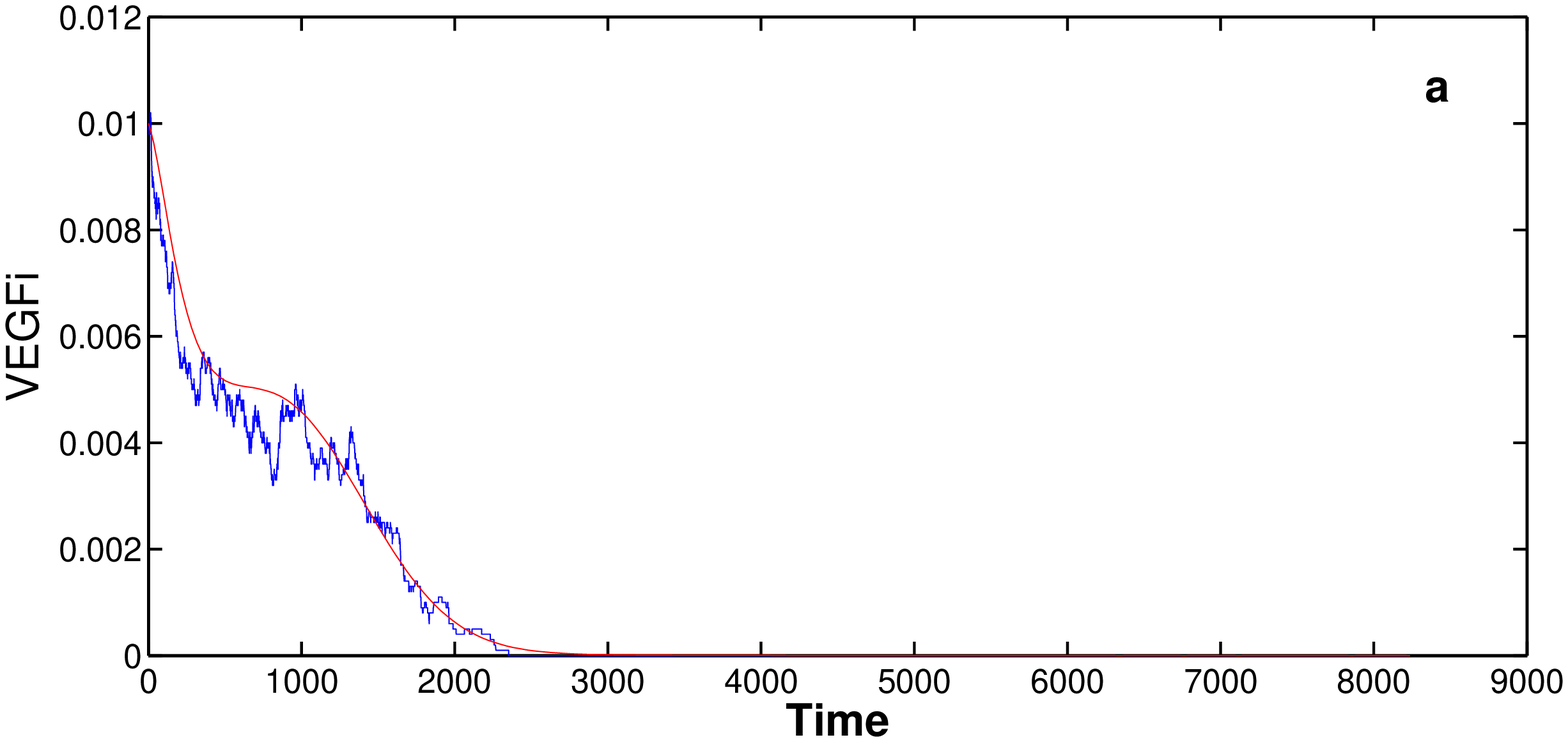} &\includegraphics[width=5.5cm, height=3.2 cm]{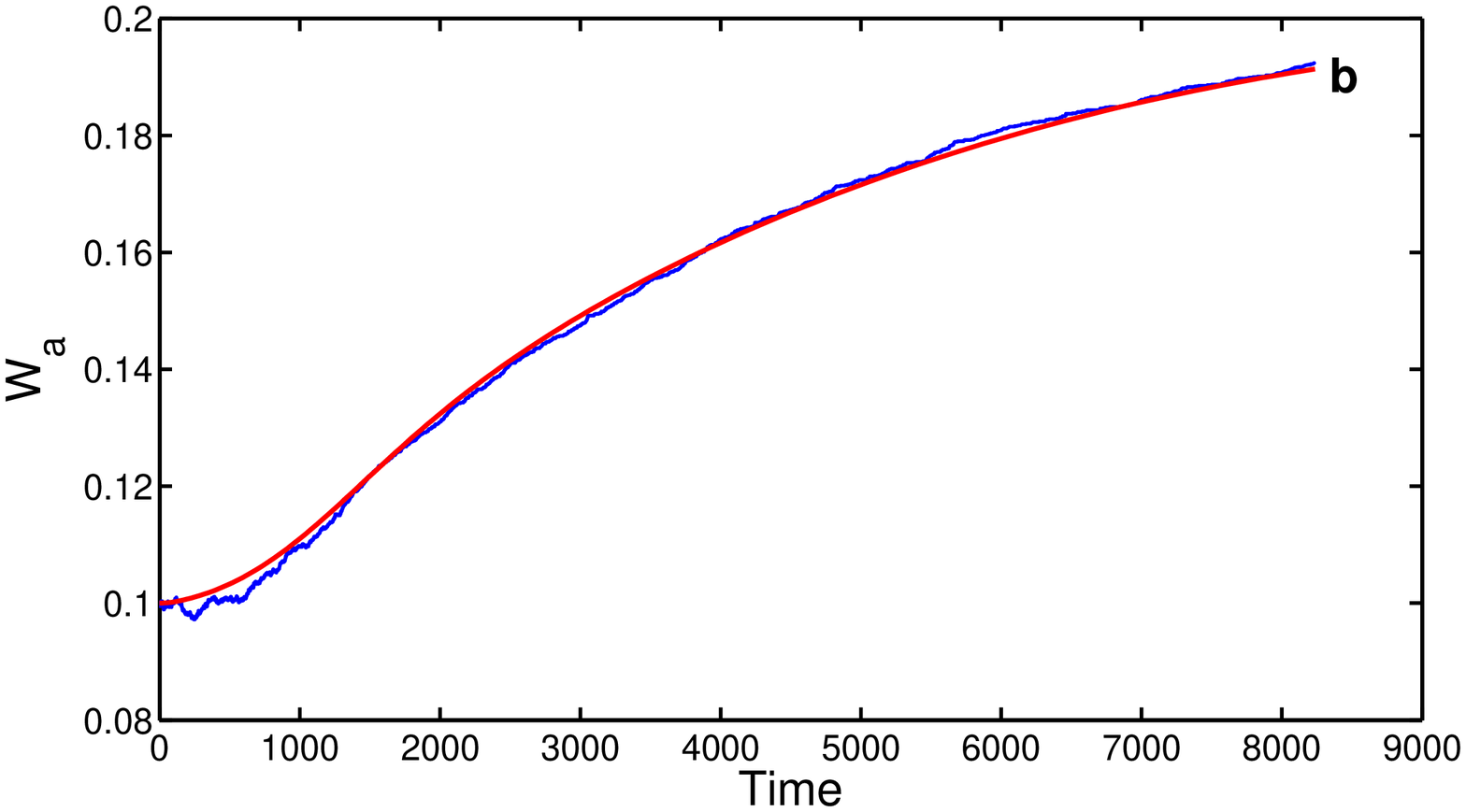} \\
\includegraphics[width=5.5cm, height=3.2 cm ]{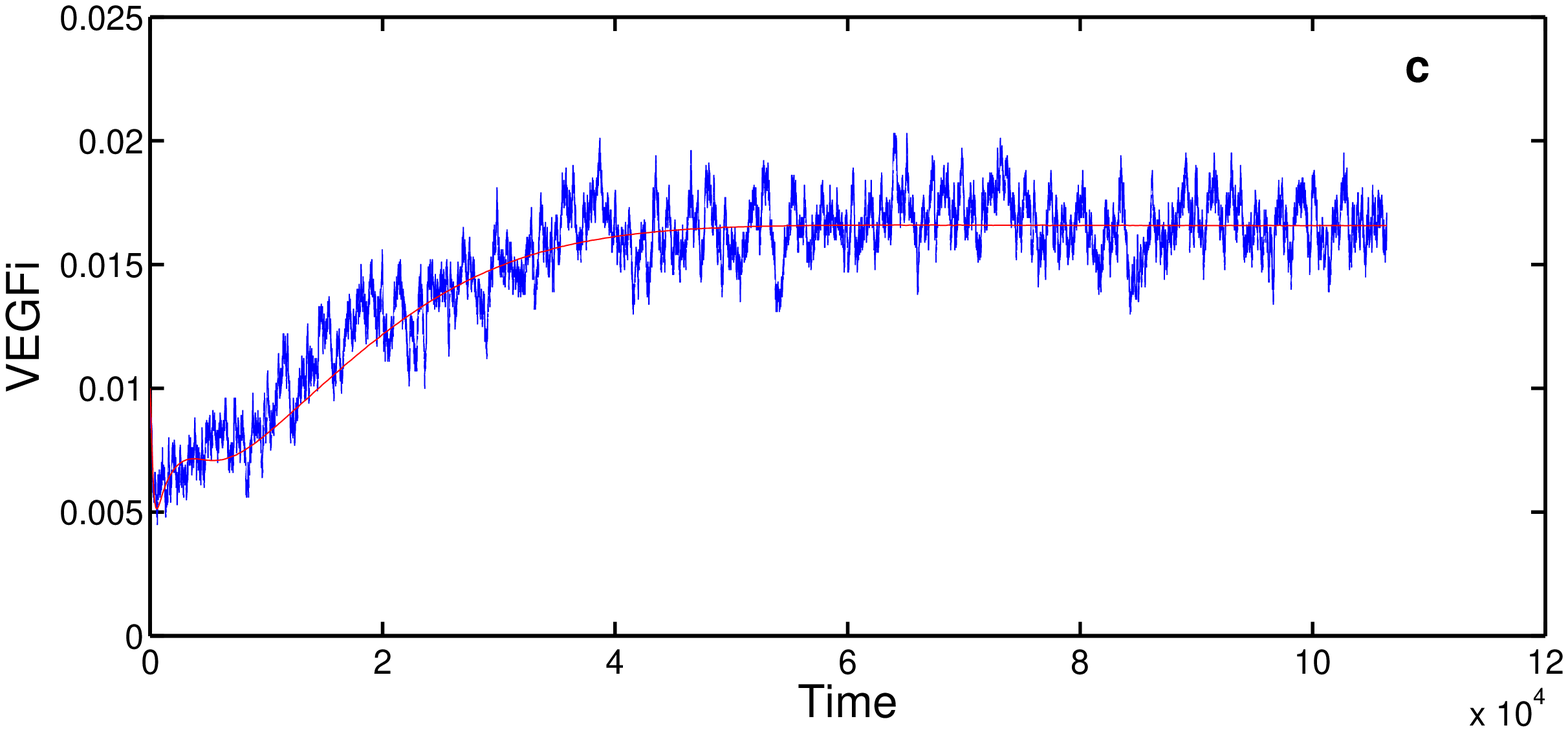} &\includegraphics[width=5.5cm, height=3.2 cm]{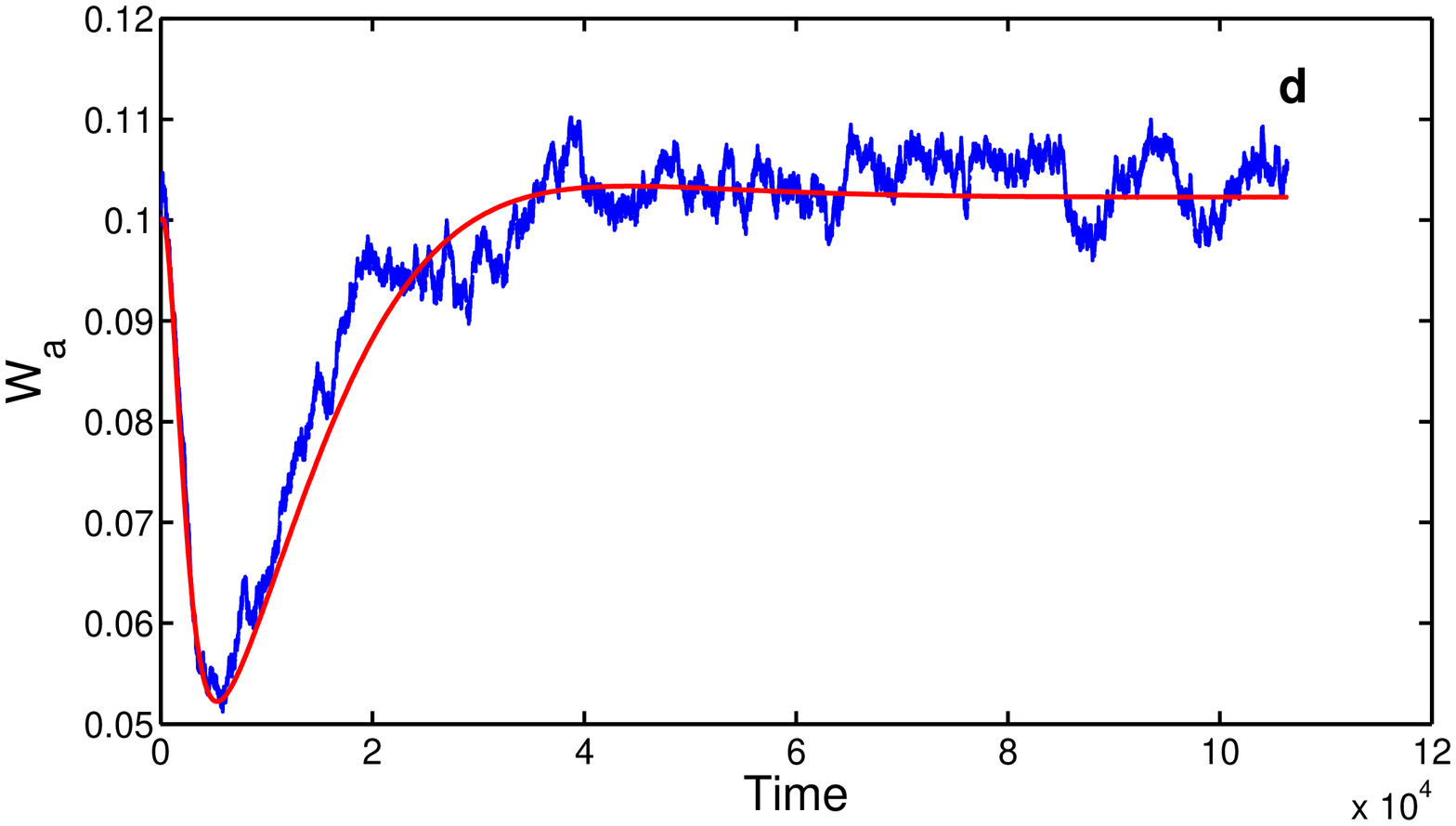} \\
\includegraphics[width=5.5cm, height=3.2 cm ]{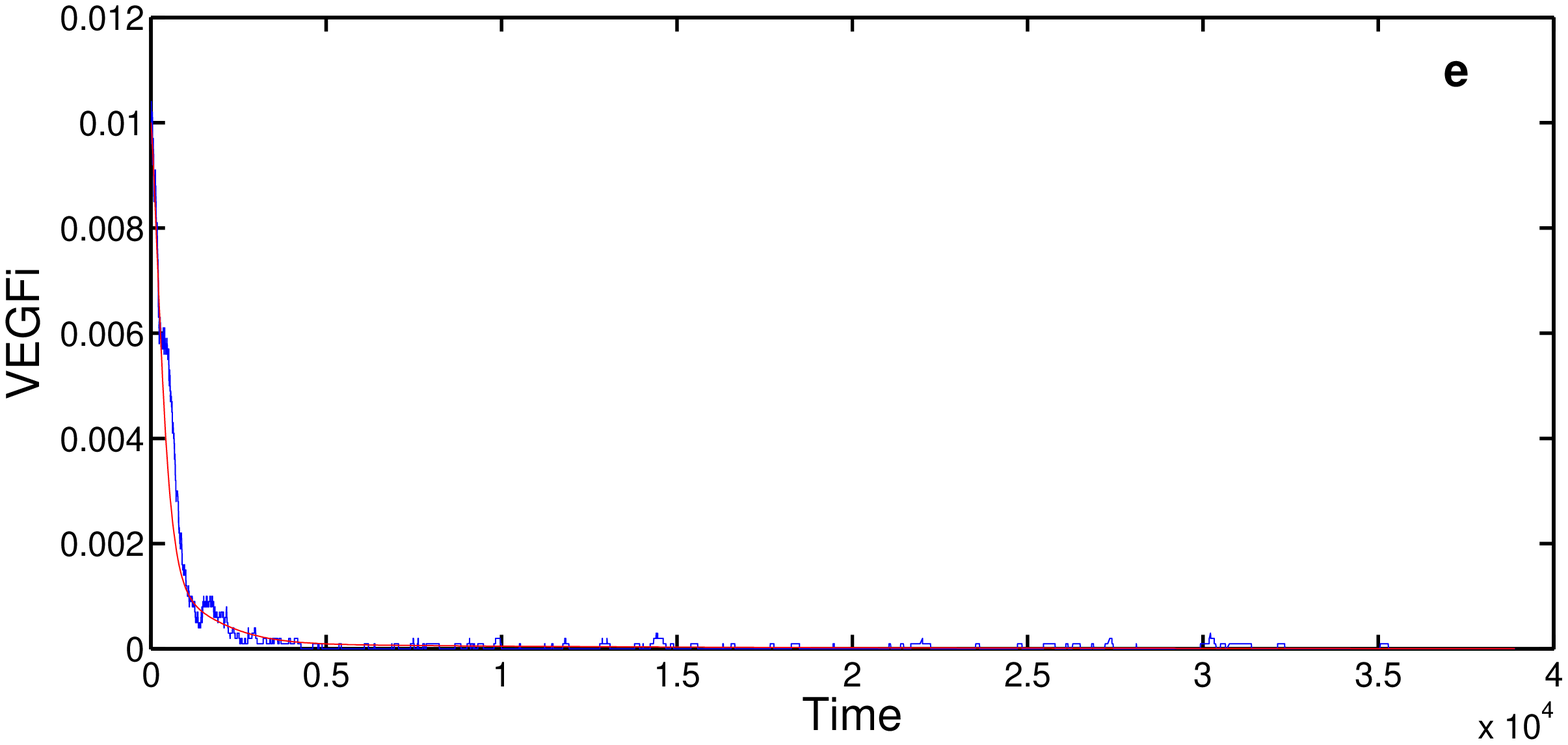} &\includegraphics[width=5.5cm, height=3.2 cm]{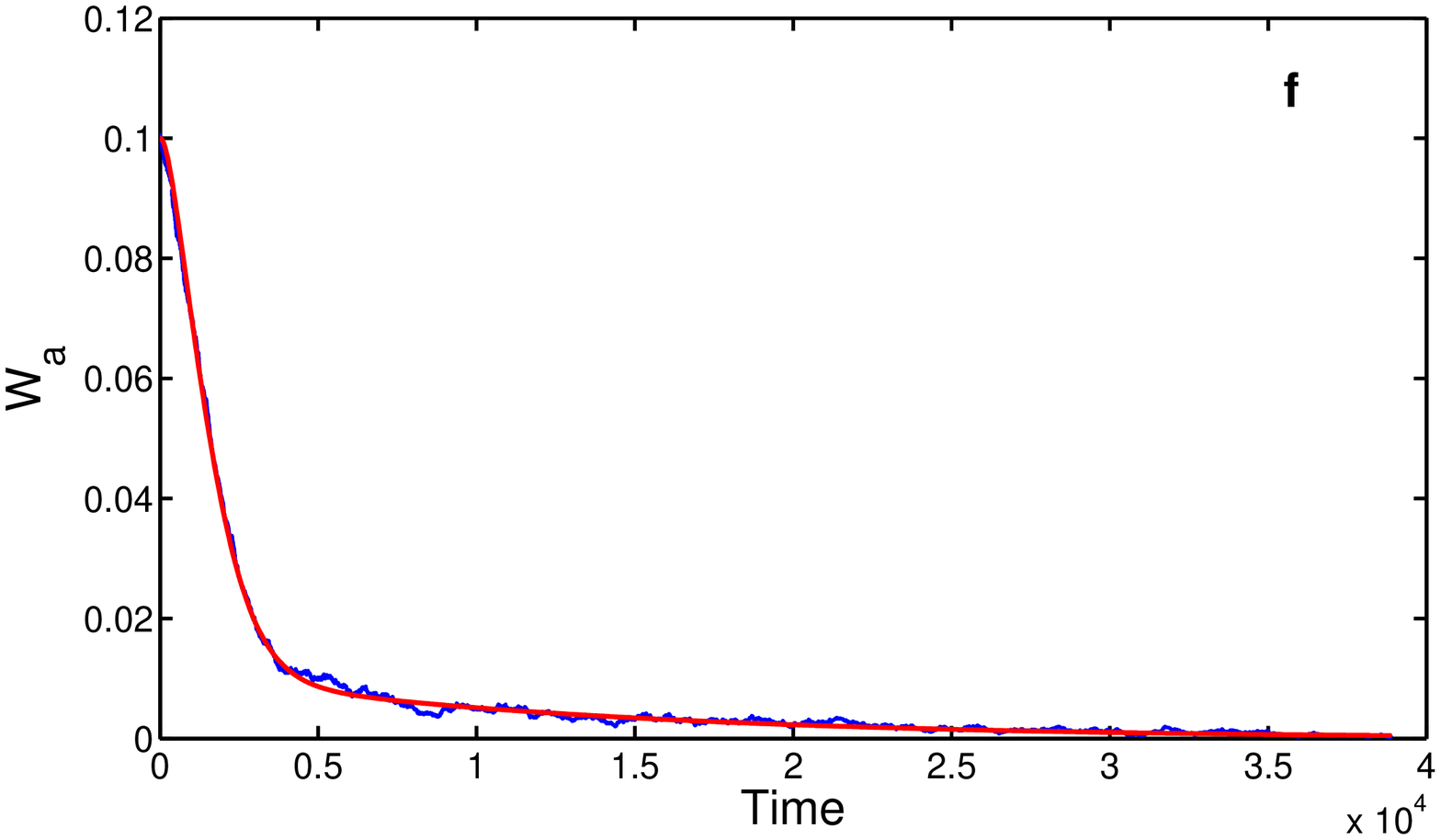}
\end{tabular}
\end{center}
\caption{Simulated  dynamics of system (\ref{eq:mf1}) -
(\ref{eq:mf8}) for different choices of the chemical parameters.
Panels (a) and  (b) report the time evolution of the concentration
of   VEGF$_i$ and  W$_a$ respectively, for $a$=0.01, $b$=0.01,
$c$=0.05, $e$=0.01, $f$=0.0005, $d$=0.0001, $g_\alpha$=0.000002,
$g_\beta$=0.0008. This is the normoxia condition, when the cell
functions, beyond the transient, without activating the hypoxia
cycle. In panel (c) and (d) the homologous quantities (resp.
VEGF$_i$ and  W$_a$) are depicted for $g_\alpha=0.0017$, while
keeping the other parameters unchanged. A residual quota of VEGF$_i$
molecules is present, which sustains the hypoxia pathway as
described in the main text. Finally, when further increasing
$g_\alpha$ ($g_\alpha=0.017$ in panels (e) and (f), respectively
referring to  the VEGF$_i$ and  W$_a$ concentrations), the cell
dies, no activity being asymptotically detected. Solid lines refer
to the integration of the mean-field dynamics (\ref{eq:mf1}) -
(\ref{eq:mf8}), while the associated irregular profiles stand for
the corresponding stochastic simulations. For all the plots
$N=10000$, the initial number of O$_2$, W, W$^*$ and HIF-$\alpha$ is
$1000$, that of HIF-$\beta$, VEGF$_i$, VEGF$_0$ and E$_i$ is equal
to $100$ and, hence, the initial number of E$_0$ is $5600$.
\label{fig:stochastic_simulation}}
\end{figure}

In this section we report about a  campaign of simulations
aimed at validating the theoretical picture depicted above.
Mean-field ordinary differential equations
(\ref{eq:mf1})-(\ref{eq:mf8}) can be straightforwardly integrated
via a standard Runge-Kutta scheme. In Fig.
\ref{fig:stochastic_simulation} the time evolution of VEGF$_i$ and
active hydroxylases is reported (solid lines) for different values
of the parameters. Panels (a) and (b) refer to a choice of the
parameters that yield to the  normoxia condition: Once the transient
has died out, the system achieves its asymptotic state where VEGF
molecules are absent. At variance, W$_a$ attains its equilibrium
concentration, different from zero. The same applies to oxygen (not
displayed here). When increasing the parameter $g_\alpha$, which
controls the HIF$-\alpha$ production rate, a transition occurs
towards what we termed the hypoxia regime, see panels (c) and  (d).
Here the HIF$-\alpha$ quota is sensibly different from zero,
implying the existence of a positive feedback reaction on the oxygen
production, via the VEGF molecules. By further augmenting
$g_\alpha$, one gradually approaches the boundary of the third
allowed region, as previously outlined: When $g_\alpha$ is large
enough, all chemical concentrations drop to zero, including the
W$_a$ amount, which is here assumed to signal the cell activity.
This scenario is clearly displayed in panels (e) and  (f)  of Fig.
\ref{fig:stochastic_simulation}.

Beside  integrating the system of differential equations
(\ref{eq:mf1})-(\ref{eq:mf8}), one can turn to direct stochastic
simulations, which enables one to treat the interacting molecules as
diffusing massive entities, so respecting the finiteness of the
chemical environment. To this end, we resort to the Gillespie
algorithm: as already recalled, this is  a sophisticated Monte Carlo
scheme which enables to inspect the actual dynamics of an ensemble
made of discrete actors, whose relative interactions are specified
by an assigned set of chemical rules. The outcome of the stochastic
integration is superposed in Fig. \ref{fig:stochastic_simulation},
to the curves determined within the mean-field approximation. The
intrinsic stochastic noise materializes as a local disturbance of
the reference mean-field profile, suggesting that the discreteness
of the environment contributes modestly to the system dynamics. As
opposed to this vision, we will prove in the following that finite
size correction can be important and can eventually result in
drastic modification of the dynamical behavior, as predicted within
the mean-field scenario.


\begin{figure}[tb]
\begin {center}
\begin{tabular}{cc}
\includegraphics[width=5.5cm, height=3.5 cm ]{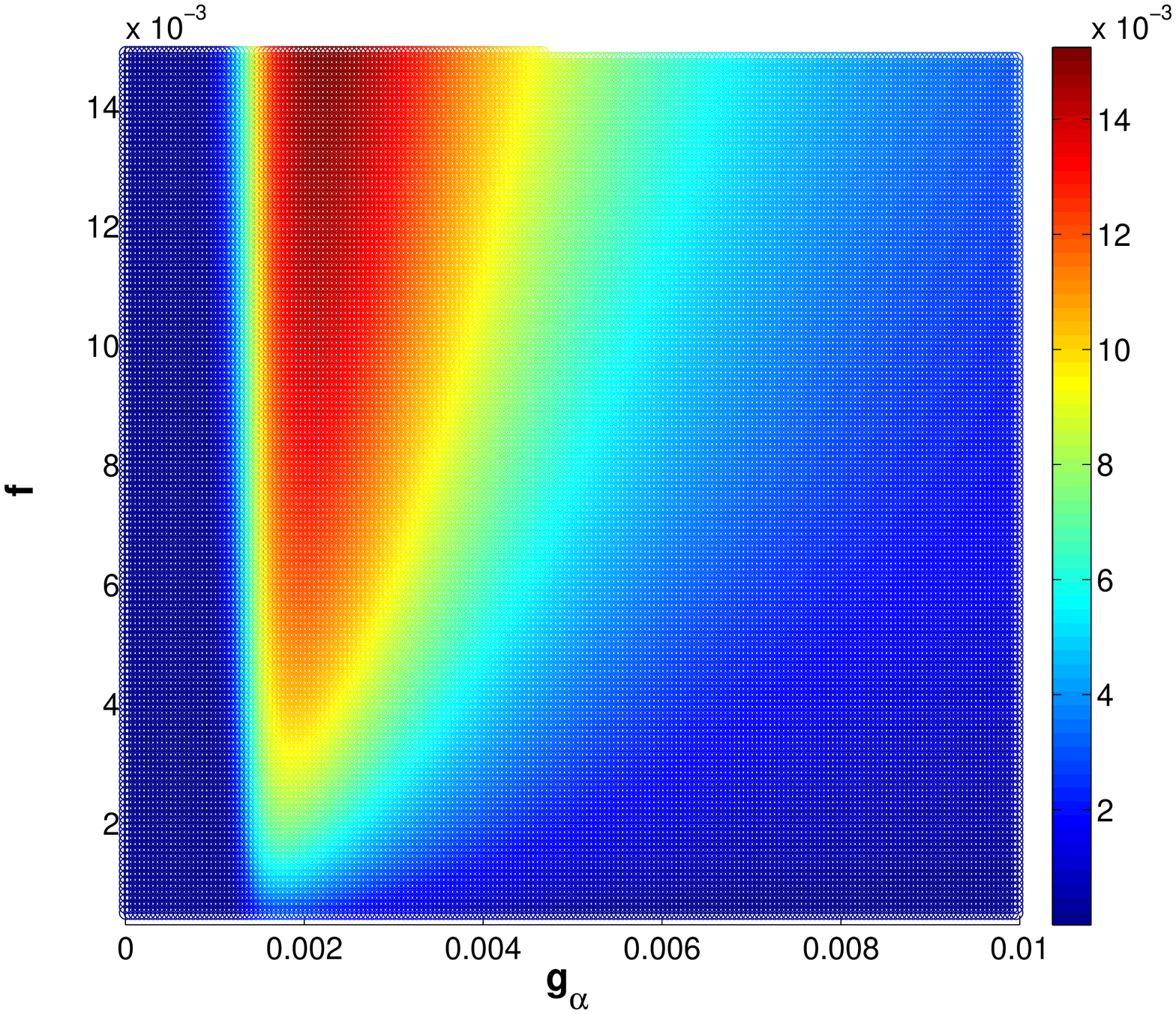} &  \includegraphics[width=5.5cm, height=3.5 cm ]{figura_taglio.eps}
\end{tabular}
\end{center}
\caption{Left panel:  The asymptotic concentration of VEGF$_i$ is
plotted for different values of $(g_\alpha,f)$. The data follows
from a direct integration of the mean-field Eqs. (\ref{eq:mf1}) -
(\ref{eq:mf8}) and are represented using the colorcode displayed in
the legend. Right panel: The concentration of VEGF$_i$ is reported
as a function of $g_\alpha$, for $f=0.005$ (the other parameters are
set as in Fig. \ref{fig:stochastic_simulation}. A double transition
is clearly displayed, the VEGF$_i$ acting as a stabilizer of the
cell dynamics for a compact range of $g_\alpha$ values. Inset a non
trivial (see text) real eigenvalue associated to the first family of
equilibrium points (normoxia), as a function of $g_\alpha$. The
eigenvalue is seen to approach zero, i.e. the threshold of stability
point, when $g_\alpha=g_\beta$. The eigenvalue is calculated via an
{\it ad hoc} semi-analytical procedure which is described in the
main body of the paper.} \label{transition_num}
\end{figure}

%
Before turning to  analyze these aspects, we provide in Fig.
\ref{transition_num} a global picture of the distinct dynamical
regimes that can be faced when tuning the $(f,g_\alpha)$ parameters:
Here the asymptotic concentration of VEGF$_i$, as calculated upon
integration of Eqs. (\ref{eq:mf1}) - (\ref{eq:mf8}) is plotted with
a colorcode. The existence of three different regions, is clearly
confirmed, the transition from normoxia (I) to hypoxia (II) being
sharper and hence more evidently appreciated. As a further
complement, we enclose a section of the scanned parameters plane, by
reporting the recorded concentration of VEGF$_i$ versus $g_\alpha$,
at fixed $f$. The double transition is shown, bearing intriguing
analogy with the phenomenon of re-entrant phase transitions.

%
%
%
%
%
\section{Dynamics in a multicellular environment: Synchronization and cooperative
effects} \label{sec:three_cells} 
\begin{figure}[tb]
\begin{center}
\includegraphics[width=8cm]{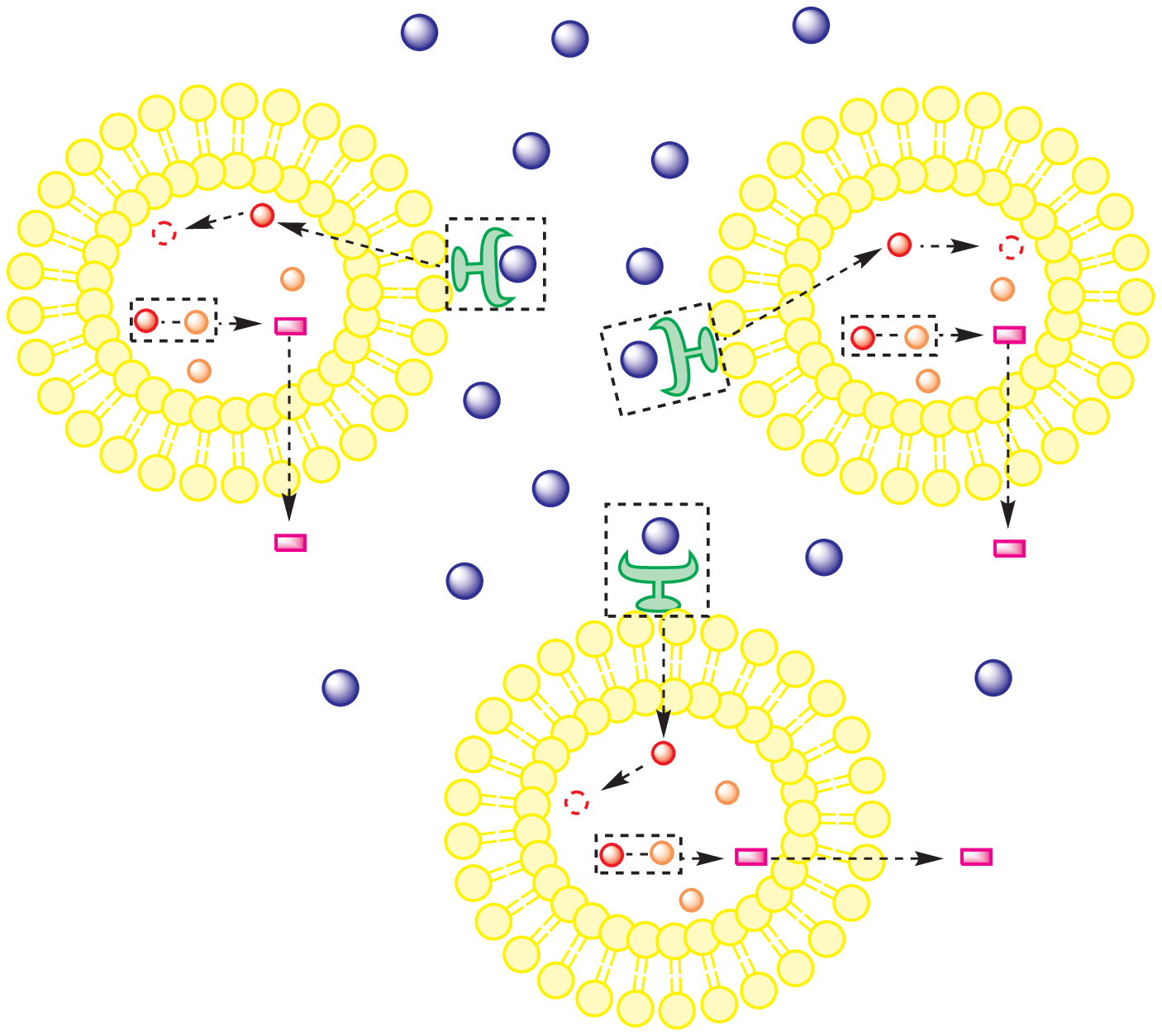}
\caption{Cartoon of the interaction scheme of  three cells. Oxygen
molecules (circles laying outside the cells' walls) are now shared,
and provides an indirect couplings among the three cells. Each cell
is then supporting a whole cycle of internal reactions, as already
exemplified in Fig. \ref{fig:cartoon}. For the symbols legend, see
caption of  Fig. \ref{fig:cartoon} } \label{fig:cartoon1}
\end{center}
\end{figure}

To shed  light onto the cooperative
effects that might eventually arise in a colony of interacting cell
units, we here consider $N_c=3$ cells obeying to the chemical
equations hypothesized above. Now the oxygen fills a shared
reservoir, from which any individual cell can draw. The cells are
hence mutually, though indirectly, communicating via the oxygen
amount. A schematic layout of the interaction network is depicted in
Fig. \ref{fig:cartoon1}. In this case we need to introduce an extra
label to specify the cell to which the molecules belong to. In
practical terms, one is forced to deal with the molecules
VEGF$_i^j$, E$_i^j$, HIF-$\alpha^j$ and HIF-$\beta^j$ with
$j=1,\ldots, N_c$. The governing chemical equations follow as a
straightforward generalization of Eqs.
(\ref{eq:chem1})-(\ref{eq:chem9}). The mean-field system calculated
from the underlying master equation is now composed by $4\times N_c
+3$ equations (not explicitly given here), the coupling being
indirectly provided by the species hosted in the outer environment
which constitutes a shared reservoir.

Working within this generalized setting, interesting cooperative
effects arise, as supported by the discrete component of the
dynamics. In the finite $N$ regime, correlations exist between the
evolving species and translate into important contributions that can
affect the system dynamics. The correlations are instead neglected
within the mean-field approximation, formally valid when the
container is filled with an infinite amount of molecular
constituents.  To demonstrate the importance of finite size
corrections, we assign to two cells a set of parameters (including
the $g_\alpha$ value) which is found to yield to the death
condition, in the mean field regime and according to the preceding
classification. The third cell is instead imagined to bear a
$g_\alpha$ parameter in the hypoxia domain (while keeping the other
parameters unchanged), as identified above. What is going to happen
to the considered system of three cells?

%
%

\begin{figure}[tb]
\begin{center}
\includegraphics[width=12cm]{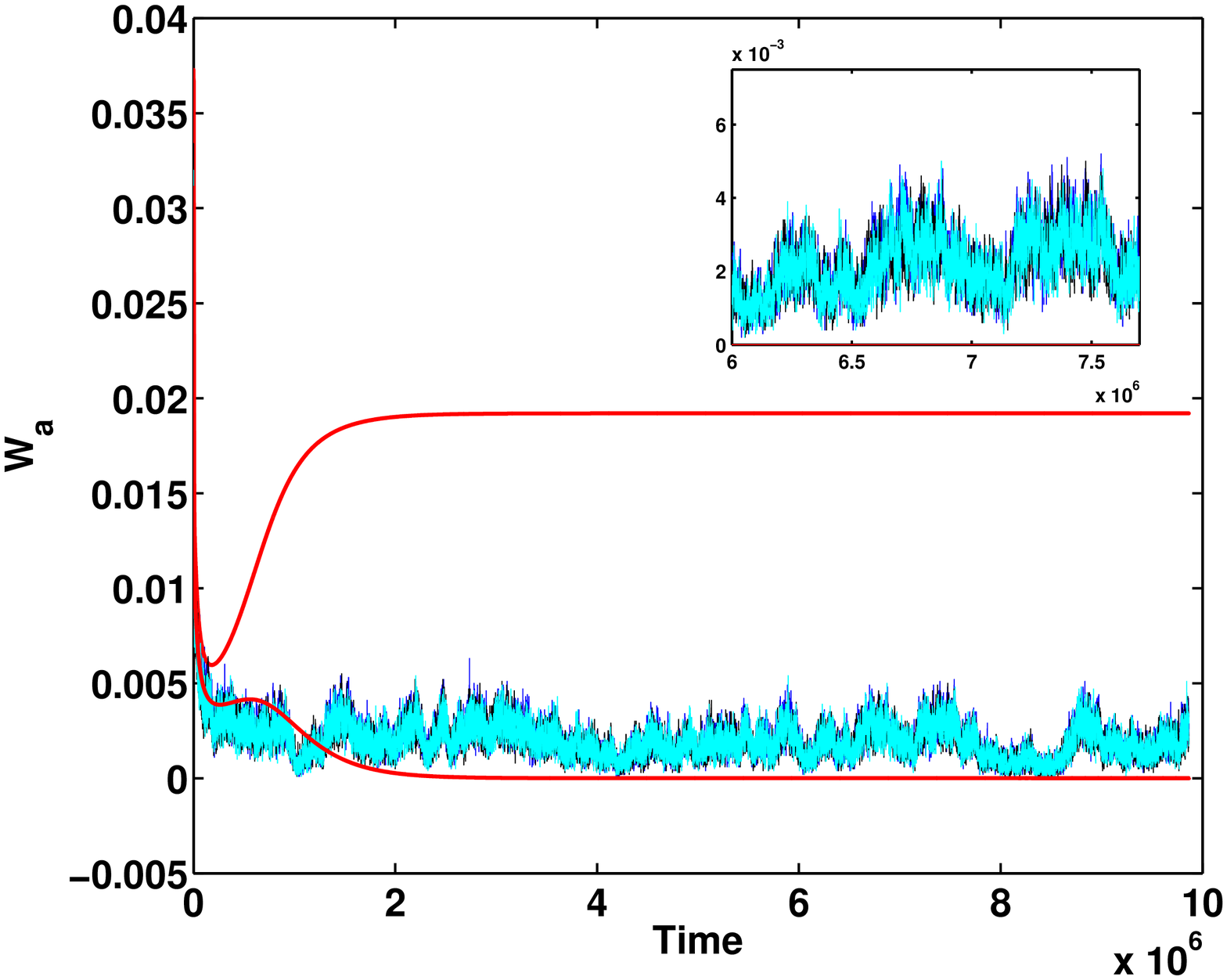}
\caption{W$_a$ evolution as a function  of time, for a system of
three cells, sharing the same oxygen reservoir. The parameters are
set so to have two cells in the mean-field region deputed to death,
while the third cell is made to function in the hypoxia domain, via
a appropriate mutation of the parameter $g_\alpha$. All other
parameters are identical for the three cells. Solid lines (red
online) represent the mean-field dynamics, while the irregular
curves (cyan online) stand for the stochastic simulations. Inset: a
zoom of the dynamical evolution is represented.
\label{fig:synchronization}}
\end{center}
\end{figure}

In Fig. \ref{fig:synchronization} the evolution of hydroxylase is
reported as a function of time. The solid lines (red online) stand
for the mean-field dynamics, obtained by numerical integration of
the generalized set of ordinary differential equations which
accounts for the three inspected cells. Two cells are destined to
die (no residual W$_a$), while the third survives (W$_a$ are
eventually present). Interestingly the stochastic dynamics is
peculiarly different, as displayed in Fig. \ref{fig:synchronization}
(wiggling curves, cyan online). As resulting from a cooperative
mechanism, ultimately driven by the finiteness of the simulated
medium, and so not captured within the corresponding continuous
description, the two cells destined to death are rescued by the
third cell. This latter can sustain the angiogenesis process and so
integrate the oxygen quota which is necessary for the correct
functioning of its nearest neighbors. The rescued cells do partially
contribute by stimulating the production of VEGF. Synchronized
oscillations are observed in the recorded  dynamics of the active
hydroxylases, as displayed in the zoomed inset of Fig.
\ref{fig:synchronization}.

In conclusion,  and based on the above findings, it is tempting to
speculate that the observed degree of cooperation turns out
particularly crucial for understanding the adaptation of  tumoral
cells in a changing environment.  Under stressing condition, it is
well known that tumoral cells can undergo mutation, changing their
specific abilities and fully exploiting their inherent flexibility.
Assume that under stressing condition the oxygen income is not
enough to guarantee the cell functioning. Then the tumoral  cells
would undergo a critical phase, which could induce punctual
mutations. In particular, imagine that following such random
adjustments, a cell enters the regime of hypoxia (in our model, by
properly tuning the corresponding $g_\alpha$ parameter). This
corresponds to reducing the HIF$-\alpha$ production rate, so
consequently freeing a quota of oxygen, otherwise employed in the
HIF$-\alpha$ degradation pathway. Therefore not only the tumoral
cell interested by the mutation can survive,  but also the cells
populating the local neighborhood (which are plausibly also
belonging to the tumoral family although not necessarily affected by
the same mutation), follows a cooperative  interaction. This effect
is here reproduced within a sound dynamical picture where finite
size effects prove  to be crucial.

%
%
%
%
%
%
\section{Conclusions}\label{sec:conclusions}
In this paper we have discussed a  simplified model to
quantitatively address the dynamical mechanisms which ultimately
underlie  the angiogenesis process. The dynamical interplay between
a limited number of molecular species is resolved and investigated
both via numerical and analytical techniques. The level of
complexity of the proposed description results from a trade-off
balance between the request of incorporating a fair amount of
biological information, and the need of making the problem
analytically tractable. Within the proposed formulation, the cell
can follow different routes, depending on the specific values of the
involved chemical parameters. Distinct macroscopic regimes are here
discussed and associated to normoxia, hypoxia and death conditions
respectively. The rate of HIF$-\alpha$ production, as well as the
parameter that controls the ability of VEGF to call for new oxygen,
are shown to play a crucial role in determining the transition
between the admissible dynamical regimes. These latter are
analytically explained via a mean-field based calculation, which is
fully confirmed by direct numerical simulations. The role of
stochastic noise is also elucidated and shown to drive a degree of
synchronization between the cells belonging to a given colony. This
is a spontaneously emerging feature, stemming from the intimate
discrete nature of the molecular environment, which can potentially
relate to the attested adaptivity of localized families of tumoral
cells. In fact, it is in general believed that cancer cells can
react to the external stressing stimuli (as for the lack of oxygen)
by favoring a specific mutation, which renders them more adaptable
to the hosting environment. While this is a customarily evoked
scenario, the simultaneous occurrence of the same mutation in all
the cancer cell population appears a highly improbable event.
Cooperative survival mechanisms, as those here identified, could in
turn contribute to explain the observed robustness, via a
self-consistent dynamical argument. As an additional point, the
results of this paper can be of interest in studying the growth of a
tumor cord characterized by an aerobic-anaerobic switch triggered by
hypoxia \cite{astanin}.

\section{Acknowledgments}
The work was supported by grants from the Associazione Genitori
contro le Leucemie e Tumori Infantili Noi per Voi, Associazione
Italiana per la Ricerca sul Cancro, Istituto Toscano Tumori to AA,
Ente Cassa di Risparmio di Firenze to the Dipartimento di Patologia
e Oncologia Sperimentali, Universit\`{a} degli Studi di Firenze.

\appendix
\section{Jacobian matrix}
The Jacobian matrix associated to the system
(\ref{eq:mf1})-(\ref{eq:mf8}) reads
\begin{displaymath}\label{eq:jacobian}
\mathbf{J} = \left( \begin{array}{cccccccc}
-a \mathcal{W} & -a \mathcal{O} & 0      & 0              & 0      & 0        & f \mathcal{E}_{0} & 0 \\
-a \mathcal{W} & -a \mathcal{O} & +b \mathcal{H}_{\alpha} & +b \mathcal{W}_a           & 0      & 0        & 0   & 0 \\
+a \mathcal{W} & +a \mathcal{O} & -b \mathcal{H}_{\alpha} & -b \mathcal{W}_a           & 0      & 0        & 0   & 0 \\
0    & 0    & -b \mathcal{H}_{\alpha} & -b \mathcal{W}_a - c \mathcal{H}_{\beta}  & -c \mathcal{H}_{\alpha} & 0        & 0   & g_\alpha \\
0    & 0    & 0      & -c \mathcal{H}_{\beta}        & -c \mathcal{H}_{\alpha} & 0        & 0   & g_\beta \\
0    & 0    & 0      & +c \mathcal{H}_{\beta}        & +c \mathcal{H}_{\alpha} & -e \mathcal{E}_{0} -d  & 0   & 0 \\
0    & 0    & 0      & 0              & 0      & +e \mathcal{E}_{0}     & -d  &0\\
0 & 0 & b \mathcal{H}_{\alpha} & b \mathcal{W}_a + c
\mathcal{H}_{\beta} & c \mathcal{H}_{\alpha} & e \mathcal{E}_{0} + d
& 0 & -g_\alpha -g_\beta
\end{array} \right)
\end{displaymath}

\bibliographystyle{elsarticle-harv}
\bibliography{bibliography}
\end{document}